\documentclass{ar-1col}
\usepackage[numbers]{natbib}
\usepackage{amsmath,amssymb,bm,graphicx,enumerate}
\usepackage[left=3cm,right=3cm,top=1.5cm, bottom=1.5cm]{geometry}

\jname{Annu. Rev. Condens. Matter Phys.}
\jvol{11}
\jyear{2020}
\doi{10.1146/annurev-conmatphys-031119-050644}

\begin{document}

\markboth{H. Watanabe}{Counting Rules of Nambu-Goldstone Modes}

\title{Counting Rules of Nambu-Goldstone Modes}

\author{Haruki Watanabe
\affil{Department of Applied Physics, University of Tokyo, Tokyo 113-8656, Japan; email: haruki.watanabe@ap.t.u-tokyo.ac.jp}}

\begin{abstract}
When global continuous symmetries are spontaneously broken, there appear gapless collective excitations called Nambu--Goldstone modes (NGMs) that govern the low-energy property of the system. The application of this famous theorem ranges from high-energy, particle physics to condensed matter and atomic physics. When a symmetry breaking occurs in systems that lack the Lorentz invariance to start with, as is usually the case in condensed matter systems, the number of resulting NGMs can be fewer than that of broken symmetry generators, and the dispersion of NGMs is not necessarily linear.  In this article, we review recently established formulae for NGMs associated with broken internal symmetries that work equally for relativistic and nonrelativistic systems.  We also discuss complexities of NGMs originating from space-time symmetry breaking. Along the way we cover many illuminating examples from various context. We also present a complementary point of view from the Lieb-Schultz-Mattis theorem.
\end{abstract}

\maketitle

\tableofcontents

\section{Introduction}

The Nambu--Goldstone theorem is a powerful theorem predicting the appearance of massless particles upon spontaneous breaking of global continuous symmetries~\cite{NambuPRL,NJL,Goldstone1,Goldstone2,RevModPhys.81.1015}.  It explains, for instance, why the pion mass is so small in terms of the chiral symmetry breaking~\cite{Weinberg}.  This is one of a few rare examples of general, non-perturbative theorems applicable to a variety of systems regardless of the microscopic details.

The examples of spontaneous broken symmetries and accompanying gapless excitations are not limited to Lorentz-invariant systems~\cite{Anderson}.  For instance, the very existence of crystals around us is a result of spontaneous breaking of spatial translation symmetry, and their universal low-energy properties, such as the Debye $T^3$ law of the specific heat, can be attributed to acoustic phonons, i.e., the Nambu--Goldstone modes (NGMs) associated with the broken translation~\cite{Kittel,Ashcroft,Grosso}.  The idea of classifying phases and exploring transitions between them based on the symmetry breaking pattern encoded in order parameters is nowadays referred to as Landau's paradigms and is understood as a prerequisite for the further classification of phases from a topological perspective~\cite{SenthilAR,Weneaal3099}.

Condensed matter systems usually lack the Lorentz symmetry due to their coupling to the surrounding environment  that fixes a specific choice of the reference frame. In the absence of the restrictive constraint imposed by the Lorentz symmetry,  even the basics properties of NGMs such as the number of modes and the behavior of their dispersion relations get variety even for the same symmetry breaking pattern.  The spinwave excitation, or magnon, in ferromagnets is the classic example that deviates from the relativistic behavior. In recent years, many new examples of ``abnormal number" of NGMs were reported in different context~\cite{Miransky,Schafer,Blaschke,Ebert,Lianyi} ranging form high-density quantum chromodynamics to cold atom systems.  
 
This review article discusses recently established formulae that provide us with a coherent understanding of all of these examples based solely on the symmetry breaking pattern and an additional information on densities of globally conserved charges in ground states.  We also address spontaneous breaking of space-time symmetries and their consequences, resolving, for example, why crystals only have phonons but do not have gapless excitations for equally broken rotations.  

Throughout this review we will set $c=\hbar=1$ to simplify notations.

\section{Internal symmetries}
\label{sec:internal}

\subsection{Spontaneous symmetry breaking}
\label{sec:basics}
Let us first specify the class of symmetries we consider in this section.  Let $G$ be the symmetry group of the system of our interest. 
In the most general case, an element $g\in G$ would transform a local operator $\hat{\phi}_\alpha(\bm{x},t)$ in the following way~\footnote{Throughout this review, quantities with `hat' represent quantum mechanical operators}.
\begin{equation}
\hat{g}\hat{\phi}_\alpha(\bm{x},t)\hat{g}^\dagger=\sum_{\beta}\hat{\phi}_\beta(\bm{x}',t')[U_g(\bm{x},t)]_{\beta\alpha}.\label{transformation}
\end{equation}
Here we assume that the symmetries are \emph{global}, meaning that the unitary matrix $U_g(\bm{x},t)$ in Eq.~\ref{transformation} is independent of $\bm{x}$ or  $t$. We also restrict ourselves to \emph{internal} symmetries for which the coordinates $(\bm{x},t)$ and $(\bm{x}',t')$ in Eq.~\ref{transformation} are identical for every $g\in G$.  When $(\bm{x}',t')$ differs from $(\bm{x},t)$ for some elements of $g\in G$, the symmetry group $G$ is referred to as \emph{space-time} and it requires a more careful treatment. We will discuss such symmetries later in Sec.~\ref{sec:spacetime}.   

The symmetry of a physical ground state $|\text{GS}\rangle$ can be lower than $G$, and when this is the case we say $G$ is spontaneously broken.   An element $h\in G$ is \emph{unbroken} if $\hat{h}|\text{GS}\rangle$ and $|\text{GS}\rangle$ represent the same state, i.e., 
\begin{equation}
\hat{h}|\text{GS}\rangle=e^{i\theta_h}|\text{GS}\rangle.\label{defh}
\end{equation}
The set of unbroken symmetries forms a subgroup $H$ of $G$.  Other elements of $G$, $G\setminus H$ as a set, is said to be \emph{broken}. Spontaneous breaking of the symmetry $G$ down to $H\subset G$ results in a ground state degeneracy, and the set of degenerate ground states constitutes the coset space $G/H$.

In practice, the definition of (un)broken symmetries written in the form of Eq.~\ref{defh} is not very useful, because ket vectors in the thermodynamic limit are, strictly speaking, ill-defined. Instead one can diagnose symmetry breaking through an expectation value of local operators. To this end, let us properly choose a set of operators $\hat{\Phi}_\alpha(\bm{x},t)$, which may be composite (a product of local operators), and consider the following combination.
\begin{equation}
\delta_g\hat{\Phi}_\alpha(\bm{x},t)\equiv \hat{g}\hat{\Phi}_\alpha(\bm{x},t)\hat{g}^\dagger-\hat{\Phi}_\alpha(\bm{x},t).\label{defop1}
\end{equation}
For unbroken symmetries, Eq.~\ref{defh} immediately implies that the expectation value of $\delta_h\hat{\Phi}_\alpha(\bm{x},t)$ vanishes for any choice of $\hat{\Phi}_\alpha$. Conversely, if the ground state expectation value
\begin{equation}
\langle\delta_g\hat{\Phi}_\alpha(\bm{x},t)\rangle=\sum_\beta\langle \hat{\Phi}_\beta(\bm{x},t)\rangle (U_g-1)_{\beta\alpha}\label{defop2}\nonumber
\end{equation}
is non-vanishing for a choice of $\hat{\Phi}_\alpha$, Eq.~\ref{defh} must be violated and the symmetry $g\in G$ is spontaneously broken. The expectation value $\langle \hat{\Phi}_\alpha(\bm{x},t)\rangle$ hence distinguishes ordered phases from disordered phases and are called an \emph{order parameter}.

Just to help grasping these notations, let us briefly discuss the quantum Ising model on the cubic lattice as an example. The Hamiltonian $\hat{H}=-J\sum_{\langle\bm{x},\bm{x}'\rangle}\hat{s}_z(\bm{x})\hat{s}_z(\bm{x}')-B\sum_{\bm{x}}\hat{s}_x(\bm{x})$ is written in terms of a spin operator $\hat{s}_\alpha(\bm{x})$ on the site $\bm{x}$ that satisfies $[\hat{s}_\alpha(\bm{x}),\hat{s}_\beta(\bm{x}')]=i\epsilon_{\alpha\beta\gamma}\delta_{\bm{x},\bm{x}'}\hat{s}_\gamma(\bm{x})$. The internal symmetry of this model is $G=\{e,g\}=\mathbb{Z}_2$, where $g$  is the spin rotation about the $x$ axis by the angle $\pi$.  To diagnose the breaking of this symmetry, we set $\hat{\Phi}_\alpha=\hat{s}_z$ and $\delta_g\hat{\Phi}_\alpha(\bm{x})=-2\hat{s}_z(\bm{x})$. In the ordered phase with $\langle \hat{s}_z\rangle\neq0$, the unbroken symmetry $H$ is trivial. The coset $G/H=\mathbb{Z}_2$ in this example corresponds to the two degenerate ground states, one with $\langle \hat{s}_z\rangle>0$ and the other with $\langle \hat{s}_z\rangle<0$.

What we discussed so far applies to both \emph{discrete} and \emph{continuous} symmetries. When $G$ is continuous (i.e., when $G$ is a Lie group), the Noether theorem~\cite{Weinberg} provides a definition of the conserved current $\partial_t\hat{q}_i(\bm{x},t)+\bm{\nabla}\cdot\hat{\bm{j}}_i(\bm{x},t)=0$. In turn, the conserved charge 
\begin{equation}
\hat{Q}_i\equiv\int d^3x\,\hat{q}_i(\bm{x},t)\label{charge0}
\end{equation}
plays the role the generator of the symmetry, enabling us to represent elements of $G$ connected to the identity as
$\hat{g}=e^{i\sum_i\epsilon^i \hat{Q}_i}$.  Here and hereafter, the label $i$ distinguishes generators of $G$.  Also, to simplify the notation we assume $3+1$ dimensions in this section.
By expanding $\hat{g}$ in Eq.~\ref{defop1} in the power series of $\epsilon$ to the linear order, we see that the spontaneous breaking of the symmetry generated by $\hat{Q}_i$ can be detected by the expectation value
\begin{equation}
\langle [\hat{Q}_i,\hat{\Phi}_\alpha(\bm{x},t)]\rangle.\label{defof3}
\end{equation}

Similarly, let us take generators $\hat{Q}_\rho$ of $H$. They are unbroken and the expectation values in Eq.~\ref{defof3} all vanish.  For continuous symmetries, the coset space $G/H$ becomes a manifold whose dimension is given by \emph{the number of broken generators}
\begin{equation}
n_{\text{BG}}\equiv\text{dim}\,G/H=\text{dim}G-\text{dim}H.\label{nBG}
\end{equation}
This number agrees with the number of ``flat directions" of fluctuations of order parameters.

\subsection{Number of Nambu--Goldstone modes}
\label{subsec:counting}
When the system spontaneously breaks a continuum symmetry $G$ down to $H$ by developing an order parameter, long-wavelength fluctuations of the order parameter give rise to gapless excitations, so-called Nambu--Goldstone modes (NGMs) or Nambu--Goldstone bosons.  They are also referred to as ``massless" because the minimum of the dispersion relation $E_{\bm{k}}$ can be identified with the mass of the quasi-particle.  In order to justify the usage of momentum $\bm{k}$, we always assume that at least discrete translation symmetries remain unbroken in every spatial direction.

In Lorentz invariant systems, it has been long known that each flat direction of order-parameter fluctuations produces its own soft mode~\cite{Weinberg}. Therefore, the number of NGMs $n_{\text{NGM}}$ is always given by the number of broken generators in Eq.~\ref{nBG},
\begin{equation}
n_{\text{NGM}}=n_{\text{BG}}\quad\text{(in relativistic systems)}.\label{countingrel}
\end{equation}
When the Lorentz invariance is explicitly broken to start with, this is not always the case. Spontaneously broken symmetries still imply the appearance of NGMs~\cite{Lange1,Lange2} but the number of NGMs can be fewer than the number of broken generators in the absence of the Lorentz symmetry:
\begin{equation}
n_{\text{NGM}}\leq n_{\text{BG}}\quad\text{(in non-relativistic systems)},\nonumber
\end{equation}
which means that the one-to-one correspondence between flat directions and gapless modes can be lost. In non-relativistic systems, knowing the symmetry-breaking pattern $G\rightarrow H$ is not sufficient to predict the number of NGMs. We need an additional input about the property of ground states, which turns out to be densities of globally conserved charges~\cite{Tomas,TomasReivew}.

One of the main goal of this review is to explain the general counting rule of NGMs that applies to both relativistic and non-relativistic cases. 
It was conjectured in Ref.~\cite{PRD} and later proved in Refs.~\cite{PRL,Hidaka} that the number of NGMs is always given by
\begin{equation}
n_{\text{NGM}}=n_{\text{BG}}-\frac{1}{2}\text{rank}\rho\quad\text{(general)}.
\label{countingtot}
\end{equation}
Here, $\rho$ is a real anti-symmetric matrix defined by 
\begin{equation}
\rho_{ij}\equiv-i\frac{1}{V}\langle[\hat{Q}_i,\hat{Q}_j]\rangle=\sum_k f_{ij}^{\,\,\,k}\frac{\langle\hat{Q}_k\rangle}{V},\label{charge}
\end{equation}
where $V$ is the volume of the system  and the thermodynamic limit $V\rightarrow\infty$ is implicitly assumed. The relation $[\hat{Q}_i,\hat{Q}_j]=i\sum_k f_{ij}^{\,\,\,k}\hat{Q}_k$ is used in the second equality. (The structure constant $f_{ij}^{\,\,\,k}$ here includes possible ``central extensions"~\cite{PRD,PRX}.)  One can also express $\rho_{ij}$ using the charge density $\hat{q}_i(\bm{x},t)$ in Eq.~\ref{charge0}.  Assuming the continuous translation symmetry, we get
\begin{equation}
\rho_{ij}=-i\langle[\hat{Q}_i,\hat{q}_j(\bm{x},t)]\rangle=\sum_k f_{ij}^{\,\,\,k}\langle\hat{q}_k(\bm{x},t)\rangle.\label{charge2}
\end{equation}
The last expression suggests that $\rho_{ij}$ vanishes in Lorentz invariant systems since $\hat{q}_i$ is the temporal component of the four vector $(\hat{q}_i,\hat{\bm{j}}_i)$. Also, comparing the middle expression of Eq.~\ref{charge2} with Eq.~\ref{defof3}, we see that $\rho_{ij}\neq0$ plays the role of an order parameter characterizing the spontaneous breaking of the generator $\hat{Q}_i$.
Therefore, $\rho_{ij}$ must vanish if either $\hat{Q}_i$ or $\hat{Q}_j$ is unbroken.

The real anti-symmetric matrix $\rho$ can always be diagonalized by an orthogonal matrix into the following form,
\begin{equation}
\rho=\begin{pmatrix}
\begin{array}{ccccccc|c}
0&\lambda_1&&&&&&\\
-\lambda_1&0&&&&&&\\
&&0&\lambda_2&&&&\\
&&-\lambda_2&0&&&&\\
&&&&\ddots&&&\\
&&&&&0&\lambda_m&\\
&&&&&-\lambda_m&0&\\\hline
&&&&&&&O\\
\end{array}
\end{pmatrix}.
\label{OrhoO}
\end{equation}
Here, $\lambda_\ell$'s ($\ell=1,\ldots,m$) are assumed to be nonzero and all blank entries are 0.  Since the matrix rank is basis-independent, we get $\text{rank}\rho=2m$  so that the right-hand side of Eq.~\ref{countingtot} is guaranteed to be an integer.

In fact, it is possible to formulate a finer counting rule of NGMs. To this end we classify NGMs into two types, type A and B~\cite{PRD,PRL}.  In the basis choice of Eq.~\ref{OrhoO}, type-B modes are those associated with pairs of broken generators $(\hat{Q}_{2\ell-1},\hat{Q}_{2\ell})$ ($\ell=1,2,\cdots,m$). As it becomes clear shortly, each pair produces only one gapless mode that normally has a quadratic dispersion $E_{\bm{k}}\propto k^2$ ($k\equiv|\bm{k}|$) in the long-wavelength limit, unless one fine-tunes parameters.   Type-A modes, on the other hand, correspond to remaining broken generators. Each of such generator produces one type-A mode that tends be linearly dispersive ($E_{\bm{k}}\propto k$) for small $k$.  By definition, the number of each type of NGMs is given by
\begin{equation}
n_{\text{A}}=n_{\text{BG}}-\text{rank}\rho,\quad n_{\text{B}}=\frac{1}{2}\text{rank}\rho,\label{countingAB}
\end{equation}
which, of course, satisfies $n_{\text{A}}+n_{\text{B}}=n_{\text{NGM}}$.  In particular, NGMs in Lorentz invariant systems are all type A.

\subsection{Historical review}
Let us review the preceding works that led to the general formulae in Eqs.~\ref{countingtot} and \ref{countingAB}.  In 1976, Nielsen and Chadha~\cite{NC} presented a general counting rule of NGMs valid either with or without relativistic invariance. They divided NGMs into two classes, type I and type II, based on the behavior of their dispersion relation. Type I and II modes, respectively, have an energy dispersion proportional to an odd and even power of its momentum in the limit of long wavelengths. By examining the analytic property of the correlation function in Eq.~\ref{defof3}, they showed that
\begin{equation}
n_{\text{I}}+2n_{\text{II}}\geq n_{\text{BG}},\label{countingNC}
\end{equation}
where $n_{\text{I}}$ and $n_{\text{II}}$ are the number of the type I and type II modes.  If $n_{\text{I}}$ and $n_{\text{II}}$ are replaced by $n_{\text{A}}$ and $n_{\text{B}}$, the inequality is always saturated:
\begin{equation}
n_{\text{A}}+2n_{\text{B}}= n_{\text{BG}}.\nonumber
\end{equation}
In fact, there exists a trivial noninteracting example in which the equality of Eq.~\ref{countingNC} does not hold under a fine-tuning of parameters (see appendix A of Ref.~\cite{PRD}).

In 2001, Sch\"afer {\it et al}~\cite{Schafer} proved a theorem, stating that $n_{\text{NGM}}=n_{\text{BG}}$ as long as $\langle[\hat{Q}_i,\hat{Q}_j]\rangle=0$ for any pair of $i$ and $j$, via a simple analysis on the linear dependence of states $\{\hat{Q}_i|\text{GS}\rangle\}$.  This is indeed the spacial case ($\text{rank}\rho=0$) of Eq.~\ref{countingtot}.  They also identified the mechanism of the reduction of $n_{\text{NGM}}$ in a concrete field-theoretical model that we review in Sec.~\ref{subsec:Kaon} --- they found a term in the Lagrangian
 that makes fluctuations into two orthogonal directions canonically conjugate to each other. We will see in Sec.~\ref{subsec:effective} that this is actually the general case when the reduction of $n_{\text{NGM}}$ occurs for internal symmetries.

In 2004~\footnote{The received date on the published paper is Dec. 26 of 2002.},  Nambu~\cite{Nambu2004} made similar but insightful observations (without a formal proof): (i) The relation $\langle[\hat{Q}_i,\hat{Q}_j]\rangle\neq0$ makes two would-be NGMs corresponding to $\hat{Q}_i$ and $\hat{Q}_j$ canonically conjugate to each other; (ii) The number $n_{\text{NGM}}$ reduces by one for every such pair; and (iii) The NGMs associated with $\langle[\hat{Q}_i,\hat{Q}_j]\rangle\neq0$ has a quadratic dispersion.  One can see that these are essentially the same statements as in the above formulae, given the existence of the basis in which the matrix $\rho$ takes the form of Eq.~\ref{OrhoO}.

With these preceding studies, the value of the general formulae in Eqs.~\ref{countingtot} and \ref{countingAB} lies in the precise formulation to the above intuitive, empirical understanding. The question is how to verify them on the general ground and we review a proof in Sec.~\ref{subsec:effective}.

\subsection{Examples}
Before going into the proof, let us see how the formulae work by reviewing pedagogical examples.

\subsubsection{Superfluid}
\label{subsec:superfluid}
\begin{figure}
	\begin{center}
		\includegraphics[width=0.6\columnwidth]{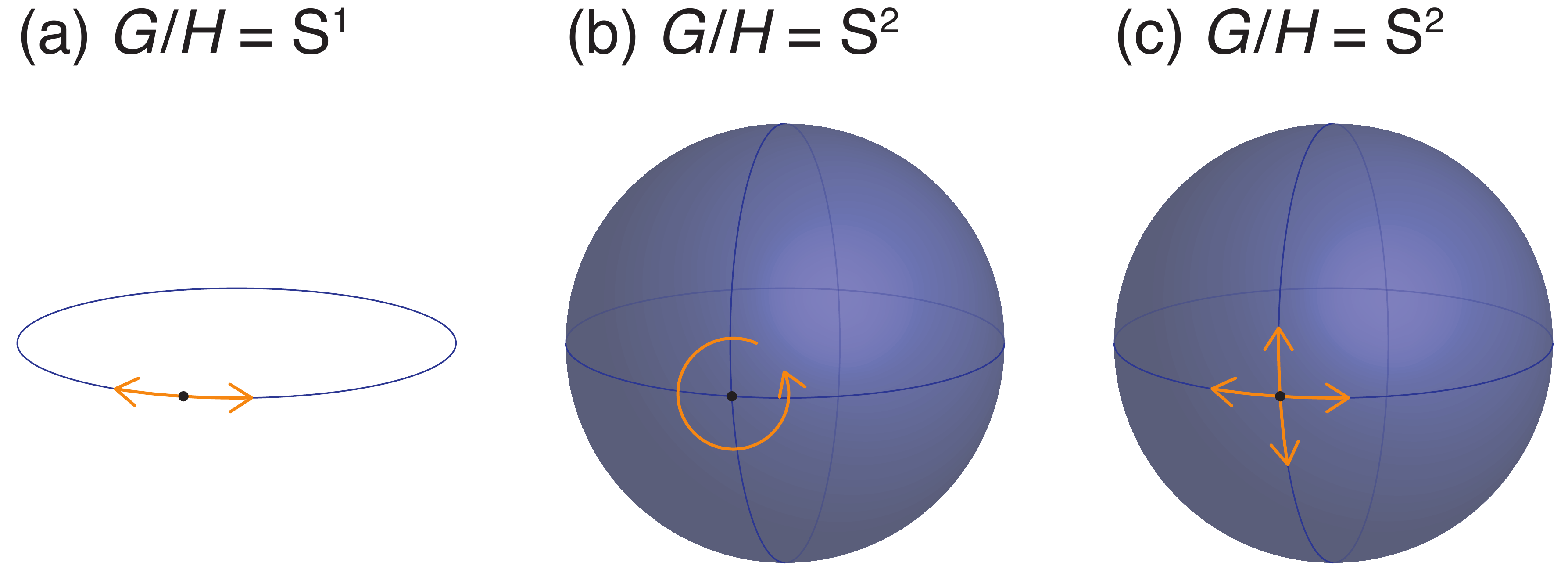}		
		\caption{\label{fig}  Illustration of the coset space $G/H$. The black dot stands for the chosen ground state and orange arrows represent fluctuations  corresponding to NGMs. (a) $G/H=\text{S}^1$ for superfluids, which may be viewed as the bottom of the ``Mexican hat" potential.
		(b) $G/H=\text{S}^2$ for ferromagnets. Two coordinates together form a single precession mode.
		(c) $G/H=\text{S}^2$ for antiferromagnets. Fluctuations into two directions are independent.
		}
	\end{center}
\end{figure}

The simplest example of spontaneous breaking of internal symmetries in condensed matter physics occurs in Bose-Einstein condensates of interacting bosons.  The original symmetry $G=\text{U}(1)$, underlying the particle-number conservation, is spontaneously broken to the trivial subgroup $H=\{e\}$.  The coset space is $G/H$ is the ring $\text{S}^1$ corresponding to the choice of the $\text{U}(1)$ phase of the macroscopic order parameter [Figure~\ref{fig}(a)]. Since there is only one generator in this example, the matrix $\rho$ always vanishes and $(n_{\text{A}},n_{\text{B}})=(1,0)$. Namely, there is no fundamental difference between the relativistic case and non-relativistic case as far as the NGM in the long-wavelength limit is concerned. The predicted type A mode is the superfluid phonon that has a linear dispersion in the long-wavelength limit (see, for example, Refs.~\cite{Nozieres,PethickSmith,ueda2010fundamentals}).  

\subsubsection{Heisenberg spin model}
\label{subsec:Heisenberg}
As a classic example that realizes $n_{\text{NGM}}< n_{\text{BG}}$, let us discuss the Heisenberg spin model on the cubic lattice.
\begin{equation}
\hat{H}\equiv J\sum_{\langle \bm{x},\bm{x}'\rangle} \hat{\bm{s}}(\bm{x}) \cdot \hat{\bm{s}}(\bm{x}').\nonumber
\end{equation}
The sum is over nearest neighbours.  This model has the full spin rotation symmetry $G=\text{SO}(3)$ generated by $\hat{Q}_i=\sum_{\bm{x}} \hat{s}_i(\bm{x})$ ($i=x,y,z$).   

When $J<0$, the ground state develops a ferromagnetic order, which we set $\langle \hat{\bm{s}}(\bm{x})\rangle=(0,0,m_z)^T$ with $m_z>0$.  Since $\rho_{xy}=-\rho_{yx}=m_z$ and other components of $\rho$ vanish, we see that both $\hat{Q}_x$ and $\hat{Q}_y$ are spontaneously broken ($n_{\text{BG}}=2$) and that $\text{rank}\rho=2$. The unbroken symmetry is $H=\text{SO}(2)$ generated by $\hat{Q}_z$.  The coset space $G/H$ is the two sphere $\text{S}^2$ [Figure~\ref{fig}(b)] in accordance with the fact that the magnetization could point in arbitrary direction.  Correspondingly, there are two flat directions for fluctuations of the ferromagnetic order parameter.   However, as it can be seen by the linearized spin-wave theory~\cite{Grosso,Auerbach}, there is only one gapless mode of the precession type, which has a quadratic dispersion. This is in agreement with $(n_{\text{A}},n_{\text{B}})=(0,1)$ indicated by Eq.~\ref{countingAB}.

When $J>0$, in contrast, the ground state has the vanishing magnetization but instead develops a N\'eel order, $\langle \hat{\bm{s}}(\bm{x})\rangle=e^{i\bm{k}\cdot\bm{x}}(0,0,n_z)^T$ ($n_z>0$, $\bm{k}=\frac{\pi}{a}(1,1,1)$, and $a$ is the lattice constant). This order also breaks the spin rotation symmetry $G=\text{SO}(3)$ down to the subgroup $H=\text{SO}(2)$ ($n_{\text{BG}}=2$).  Although the symmetry breaking pattern is identical to the ferromagnetic case as far as the internal symmetry is concerned, this time all components of $\rho$ vanishes and the relativistic result $(n_{\text{A}},n_{\text{B}})=(2,0)$ is recovered. Thus there exist two linear modes  and  $n_{\text{NGM}}=2$ [Figure~\ref{fig}(c)].  In this example, the value $\rho=0$ is ``protected" by an unbroken symmetry of the time-reversal operation followed by the unit translation.  

The effective Lagrangian treatment of NGMs in ferromagnets and antiferromagnets can be found, for example, in Refs.~\cite{Leutwyler,Burgess,Fradkin}.

\subsubsection{Kaon condensation in QCD}
\label{subsec:Kaon}
The above spin model may look too unfamiliar to people with high-energy background.  Thus let us discuss an illuminating example of a scalar field theory, describing the Kaon condensation in the color-flavor locked phase of quantum chromodynamics~\cite{Miransky,Schafer}. The Lagrangian density for  a two-component complex scalar field $\phi=(\phi_1,\phi_2)^T$ reads
\begin{equation}
\mathcal{L}\equiv \sum_\nu(\partial^\nu+iA^\nu)\phi^\dagger (\partial_\nu-iA_\nu)\phi-\mathcal{V}(\phi^\dagger\phi),\nonumber
\end{equation}
where $\mathcal{V}(\phi^\dagger\phi)\equiv m^2\phi^\dagger\phi+\lambda(\phi^\dagger\phi)^2$ is a $\phi^4$-potential, $x^\nu$ ($\nu=0,1,2,3$) corresponds to $(t,x,y,z)$, and $A_\nu$ is a background $\text{U}(1)$ gauge field which we set $A_0=\mu>0$ and $\bm{A}=\bm{0}$.  Clearly, the Lorentz invariance is explicitly broken by the chemical potential $\mu\neq0$.  The internal symmetry of the model is $G=\text{U}(2)$ that has four generators $\hat{Q}_\nu$.

The (classical) expectation value $\phi_0$ is found by minimizing the potential $\mathcal{V}_\mu(\phi^\dagger\phi)\equiv \mathcal{V}(\phi^\dagger\phi)-\mu^2\phi^\dagger\phi$ modified by the chemical potential.
When $\mu>m$, $\phi_0$ can be set $(v,0)^T$ with $v>0$, which breaks the symmetry $G=\text{U}(2)$ into a subgroup $H=\text{U}(1)$ generated by $\hat{Q}_0-\hat{Q}_z$. The coset space $G/H$ is the three sphere $\text{S}^3$ and $n_{\text{BG}}=3$.  At the tree-level approximation, the expectation value of the Noether charge is given by
\begin{equation}
\langle \hat{Q}_\nu\rangle\simeq\int d^3x\phi_0^\dagger(i\partial_t+\mu)\sigma_\nu\phi_0+\text{c.c.}=2\mu v^2V\delta_{\nu,z},\nonumber
\end{equation}
where $\sigma_\nu$'s are Pauli matrices. The nonzero components of $\rho$ are $\rho_{xy}=-\rho_{yx}=4\mu v^2\neq0$ and $\text{rank}\rho=2$. 

Now, let us introduce a small fluctuation $\pi^a$ for each broken generator $\hat{Q}_a$ ($a=x,y,z$). Plugging $\phi=\phi_0+v(i\pi^z,\pi^y-i\pi^x)$ into the Lagrangian and expand it to the quadratic order in $\pi^a$, one finds a term $2\mu v^2(\partial_t{\pi}^x\pi^y -\partial_t{\pi}^y\pi^x)$~\cite{Schafer}. As a result, $\pi^x$ and $\pi^y$ are paired up to be a single quadratic mode in the low-energy limit, while $\pi^z$ produces a linearly dispersive mode~\cite{Miransky,Schafer}.  This is consistent with the prediction $(n_{\text{A}},n_{\text{B}})=(1,1)$ based of the general counting rules.

This result should be compared to the standard Lorentz-invariant case with $\mu=0$ and $m^2<0$. The same form of the classical expectation value $\phi_0=(v',0)^T$ implies the breaking of the three generators ($n_{\text{BG}}=3$). (When $\mu=0$, $G$ and $H$ are enhanced to $\text{O}(4)$ and $\text{O}(3)$ but $G/H$ remains to be $\text{S}^3$.) This time we get $\text{rank}\rho=0$ and $(n_{\text{A}},n_{\text{B}})=(3,0)$, i.e., there are three gapless modes with the relativistic dispersion $E_{\bm{k}}=k$.

\subsubsection{Spinor BEC}
Cold atomic systems with spin degrees of freedom offer a nice playground of variety of intriguing symmetry breaking patterns. For example, the $F=1$ spinor Bose-Einstein condensate can realize the ferromagnetic phase that breaks $G=\text{U}(1)\times\text{SO}(3)$ down to $H=\text{U}(1)'$ (the prime indicates that the two $\text{U}(1)$ factors in $G$ and $H$ are different)~\cite{Ho,Ohmi}. The nonzero magnetization implies $\text{rank}\rho=2$, and the counting rule predicts $(n_{\text{A}},n_{\text{B}})=(1,1)$.  Indeed, the superfluid phonon with a linear dispersion relation and the spin-wave with a quadratic dispersion co-exist in this phase.  The dynamics of macroscopic order in Bose-Einstein condensates can be described by the Bogoliubov theory (also called the Gross-Pitaevskii theory), which formally looks very similar to the above model of Kaon condensation. NGMs in this class of models were studied in Ref.~\cite{TAKAHASHI2015101}. As there already exist several nice reviews \cite{UedaAR,Ueda1,Ueda2,Cazalilla}, here we avoid repeating the details.  We still add that the spin wave dispersion in the ferromagnetic phase has been measured in a recent experiment of ${}^{87}\text{Rb}$~\cite{EdMarti}.

\subsection{Effective Lagrangian}
\label{subsec:effective}
Now let us overview the derivation of general formulae discussed in Sec.~\ref{subsec:counting} in the effective Lagrangian approach~\cite{PRL,PRX,Antonio}. (For a general review of effective field theories, see Refs.~\cite{ARN1,ARN2,ARN3}.)  The formulae were also verified independently in Mori's projection operator formalism~\cite{Hidaka} and have been extended to systems at finite temperatures~\cite{HayataHidaka} or out of equilibrium~\cite{Minami}.  

The effective Lagrangian is designed to capture the low-energy, long distance fluctuations of the macroscopic order of the system.  The only input to the effective theory is the symmetry breaking pattern $G\rightarrow H$. One constructs the most general Lagrangian by including all terms that respect assumed symmetries~\cite{CWZ,CCSWZ,WeinbergPA}.  To control the number of terms in the Lagrangian, we make two simplifications. (i) Only fluctuations into the direction of $G/H$ are taken into account. In other words, fluctuations of the amplitude of the order parameters, giving birth to the Higgs modes~\cite{Varma}, are neglected.  The effective theory is thus the non-linear sigma model whose target space is $G/H$. (ii) Since the focus is on the the low-energy, long-wavelength dynamics, the derivative expansion that arrange terms in the power series of derivatives is employed. In the absence of Lorentz invariance, the time and space derivatives have to be counted separately.  
For simplicity we assume the spatial rotation symmetry; otherwise the power-counting of $\partial_x, \partial_y, \cdots$ should be all independent. We also assume first that there are no low energy degrees other than NGMs in the system. When they do exist they can be added later. 

The general structure of the effective Lagrangian for non-relativistic systems was studied in details in the pioneering work of Leutwyler~\cite{Leutwyler,Leutwyler2}.  The advantage of using the Lagrangian formalism is that commutation relations among fields upon the quantization are an \emph{output}, not an input, of the theory in contrast to the Hamiltonian formalism as in Ref.~\cite{Halperin}.

As a warm-up, let us discuss the effective Lagrangian describing the superfluid phase of interacting bosons (Sec.~\ref{subsec:superfluid}).
The order parameter of the condensate can be written as $\Phi(\bm{x},t)=\sqrt{n(\bm{x},t)} e^{i\theta(\bm{x},t)}$. Here the field $\theta$ parameterizes the coset space $G/H=\text{S}^1$ and transforms nonlinearly under the broken $\text{U}(1)$ symmetry $\hat{g}=e^{i\epsilon\hat{Q}}$,
\begin{equation}
\theta'(\bm{x},t)=\theta(\bm{x},t)+\epsilon.
\label{theta}
\end{equation}
The effective Lagrangian, invariant under the $\text{U}(1)$ symmetry, is thus written in terms of derivatives of $\theta$:
\begin{equation}
\mathcal{L}_{\text{eff}}=\tfrac{1}{2}\bar{g}(\partial_t{\theta})^2-\tfrac{1}{2}g(\bm{\nabla}\theta)^2+\cdots.
\label{LSF}
\end{equation}
The dots represent terms with more derivatives, which are not important in the low-energy limit.   This Lagrangian describes a NGM with the linear dispersion relation $E_{\bm{k}}=vk$ with $v=\sqrt{g/\bar{g}}$.  

To do the same for a general symmetry breaking pattern $G\rightarrow H$, let us introduce so-called Maurer-Cartan forms that serve as building blocks of the effective Lagrangian~\cite{CWZ,CCSWZ}. Let $T_i$'s be a faithful representation of generators of $G$, satisfying $[T_i,T_j]=i\sum_k f_{ij}^{\,\,\,k}T_k$, and let $T_a$'s be broken generators. Then the Maurer-Cartan one-form $\omega_\mu^i$ ($\mu=t,x,y,z$) is defined by 
\begin{equation}
\sum_i\omega_\mu^iT_i\equiv-i U^{\dagger} \partial_\mu U,\quad U(\pi)\equiv e^{i\sum_a\pi^aT_a}.\nonumber
\end{equation}
Nambu--Goldstone fields $\pi^a$ in $U(\pi)$ can be viewed as a local coordinate of the manifold $G/H$. By the series expansion, we see that 
\begin{equation}
\omega_\mu^a=\partial_\mu \pi^a+\frac{1}{2}\sum_{b,c}f_{bc}^{\,\,\,a}\pi^b\partial_\mu \pi^c+O(\pi^3).  \label{seriesomega}
\end{equation}
For instance, the above example of the superfluid corresponds to the choice $\pi^a=\theta$ and $T_a=1$ so that $U=e^{i\theta}$ and $\omega_\mu=-i U^{\dagger} \partial_\mu U=\partial_\mu \theta$.

The transformation rule of Nambu--Goldstone fields $\pi^a$ under $g=e^{i\sum_i\epsilon_iT_i}$ is defined by 
\begin{equation}
gU(\pi)=U(\pi') h_g,\quad h_g\in H.\nonumber
\end{equation}
It follows that $\pi^a$'s transform linearly, $(\pi^a)'=\sum_b(U_h)_{ab}\pi^b$, under unbroken symmetry $h\in H$ and nonlinearly [e.g. Eq.~\ref{theta}] under broken symmetry $g\in G\setminus H$. 

To the quadratic order in derivatives, the general effective Lagrangian for $G\rightarrow H$, symmetric under the spatial translation and rotation, was found to be~\cite{PRX}
\begin{equation}
\mathcal{L}_{\text{eff}}=-\sum_{i}e_i\,\omega_t^i+\sum_{a,b}\tfrac{1}{2}\bar{g}_{ab}\,\omega_t^a\omega_t^b-\sum_{a,b}\sum_{r=x,y,z}\tfrac{1}{2}g_{ab}\,\omega_r^a\omega_r^b+\cdots.
\label{summary}
\end{equation}
Constants $e_i$, $g_{a b}$, and $\bar{g}_{a b}$ must obey the conditions $\sum_jf_{\rho i}^{\phantom{\rho i}j}e_j=0$ and $\sum_c(f_{\rho a}^{\phantom{\rho a}c}g_{c b}+f_{\rho b}^{\phantom{\rho b}c}g_{ac})=0$ (the same for $\bar{g}$). With these constraints, $\mathcal{L}_{\text{eff}}$ is fully symmetric under $G$, not only under the unbroken subgroup $H$.  As found by Ref.~\cite{Leutwyler}, the constant $e_i$ is related to the conserved charge density $e_i=\langle\hat{q}_i(\bm{x},t)\rangle$.

This effective Lagrangian is a natural generalization of the well-established result for Lorentz-invariant systems ($\eta^{\mu\nu}$ is the Minkowski metric)
\begin{equation}
\mathcal{L}_{\text{eff}}=\sum_{a,b}\sum_{\mu,\nu}\tfrac{1}{2}g_{ab}\,\eta^{\mu\nu}\omega_\mu^a\omega_{\nu}^{b}+\cdots,\nonumber
\end{equation}
which corresponds to $e_i=0$ and $\bar{g}_{ab}=g_{ab}$ in Eq.~\ref{summary}. This relativistic form reproduces the original Nambu--Goldstone theorem \ref{countingrel}.

To examine the nature of excitations described in the general effective Lagrangian, let us expand Eq.~\ref{summary} to the quadratic order in $\pi$ by plugging Eq.~\ref{seriesomega}. We get~\cite{PRL}
\begin{equation}
\mathcal{L}_{\text{eff}}\simeq\sum_{a,b}\tfrac{1}{2}\rho_{ab}\,\partial_t\pi^a\pi^b+\sum_{a,b}\tfrac{1}{2}\bar{g}_{ab}\,\partial_t\pi^a\partial_t\pi^b-\sum_{a,b}\sum_{r=x,y,z}\tfrac{1}{2}g_{ab}\,\partial_r\pi^a\partial_r\pi^b+\cdots.\label{quadratic}
\end{equation}
Assuming the form of $\rho$ in Eq.~\ref{OrhoO}, the first term implies that $\pi^{2\ell-1}$ and $\pi^{2\ell}$ are conjugate degrees of freedom, producing a type-B NGM with a quadratic dispersion ($E_{\bm{k}}=k^2/2m$ with $m\sim \rho_{ab}/g_{ab}$).  When the second term of Eq.~\ref{quadratic} is nonzero, a  ``gapped partner," whose mass gap is the order of $\rho_{ab}/\bar{g}_{ab}$~\cite{Kapustin}, may exist for each type-B mode, although such a statement for gapped modes requires further consideration on the consistency of the derivative expansion.  The remaining $\pi^a$ fields produce their own linearly dispersive modes (see Ref.~\cite{PRX} for the detailed analysis). These results imply the counting rules in Eq.~\ref{countingAB}.

The effective Lagrangian can be used not only for counting NGMs but also for analyzing interactions of NGMs.  For example, higher order terms in $\pi^a$'s in Eq.~\ref{summary}, dropped in the linearized Lagrangian in Eq.~\ref{quadratic}, describe interactions among NGMs.  Other low-energy degrees possibly existing in actual symmetry broken phases can also be included to Eq.~\ref{summary} by adding terms that respect all the assumed symmetries.

\subsection{Mermin-Wagner-Coleman theorem}
The difference between the two types of NGMs is not limited to the dispersion relation. To clarify this point, let us review the Mermin-Wagner-Coleman theorem~\cite{Hohenberg,MerminWagner,Coleman}. The standard understanding of this theorem is that it prohibits spontaneous breaking of any continuous symmetry in $1+1$ dimensions at the zero temperature (equivalently, two spatial dimensions at a finite temperature).  Intuitively, the theorem holds because the would-be gapless mode produces an uncontrolled quantum fluctuation in one spatial dimension. Using the effective Lagrangian in Eq.~\ref{quadratic}, one can readily evaluate the fluctuation due to a type-A NGM:
\begin{equation}
\langle\pi(\bm{x},t)^2\rangle\sim \int d^dkd\omega\frac{1}{\bar{g}\omega^2-gk^2}\propto \int_0^\Lambda dk k^{d-2},\notag
\end{equation}
which suffers from the infrared divergence when $d=1$. The ultraviolet cutoff $\Lambda$ is not of our interest as we are focusing on the long-wavelength physics.  Such an infrared divergence destroys the order parameter for the postulated broken symmetry.

One may think that the situation gets worse for type-B NGMs because they are softer, but this is not the case.  In fact, fluctuation caused by a type-B NGM
\begin{equation}
\langle\pi(\bm{x},t)^2\rangle\sim \int d^dkd\omega\frac{1}{i \rho\omega-gk^2}\propto \int_0^\Lambda dk k^{d-1}
\end{equation}
converges even when $d=1$. This suggests that continuous symmetries can, in fact, be spontaneously broken in $1+1$ dimensions as far as only type-B NGMs appear.  This is actually reasonable ---  according to Eq.~\ref{charge}, the order parameter for the broken generators $\hat{Q}_i$, $\hat{Q}_j$ is given by the sum of conserved charges $\rho_{ij}=\sum_kf_{ij}^{\,\,k}\langle\hat{Q}_k\rangle/V$.  Because $\hat{Q}_k$ commutes with the Hamiltonian, one can choose the physical ground state to be a simultaneous eigenstate of $\hat{Q}_k$ and the Hamiltonian. Then there would be no quantum fluctuation of the order parameter regardless of the dimensionality of the system. The trivial example is again provided by the quantum ferromagnet in which $\langle\hat{S}_z\rangle$ plays the role of order parameter. Because of this fundamental difference, sometimes these cases are excluded from the examples of spontaneous broken symmetries. (For example, see the commentary by Anderson~\cite{AndersonComment}). 

\section{Space-time symmetries}
\label{sec:spacetime}
We move on to the discussion of spontaneous breaking of space-time symmetries.  For this class of symmetries, there are several complications at different levels, and it does not seem possible to write down the most general formula in any useful manner.  Here let us instead start with examples to see what we get out of them in general.  Throughout this section we consider general spatial dimension $d$ greater than one.

\subsection{Translation symmetry}
Let us begin by spontaneously breaking of translation symmetry. In many respects the spatial translation is unique among all space-time symmetries. Most importantly, NGMs originating from the spontaneously broken translation can be treated in essentially the same way as those associated with internal symmetries.

\subsubsection{Crystals} 
\label{subsec:crystal}
Suppose that the system develops a crystalline order that breaks the continuous translation symmetry $G=\mathbb{R}^d$ down to a discrete one in every direction so that $H=\mathbb{Z}^d$.  The number of broken generators is thus $n_{\text{BG}}=d$.  To describe low-energy deformation of the crystal, we introduce the same number of displacement fields $u^a(\bm{x},t)$ that transform nonlinearly under the corresponding translation,
\begin{equation}
u^a(\bm{x}+\bm{\epsilon},t)'=u^a(\bm{x},t)+\epsilon^a.\nonumber
\end{equation}
The theory of elastic medium~\cite{Landau,CL} gives us the effective Lagrangian that describes the acoustic phonons, which reads
\begin{equation}
\mathcal{L}_{\mathrm{eff}} = \frac{1}{2}m(\partial_t\bm{u})^2-\frac{1}{2}\lambda_{abcd}\varepsilon^{ab}\varepsilon^{cd}+\cdots,\label{phononL}
\end{equation}
where $\varepsilon^{ab}\equiv\frac{1}{2}(\partial_au^b+\partial_bu^a)$ is the (linearized) strain tensor and $m$ is the mass density of the medium. Unbroken spatial symmetries of crystals reduce independent parameters in the elastic modulus tensor $\lambda_{abcd}$~\cite{Landau} (also known as the elastic constant tensor~\cite{CL}). The Lagrangian in Eq.~\ref{phononL} predicts the existence of $d$ linearly dispersive modes -- one longitudinal and $(d-1)$ transverse in the isotropic case. The number of NGMs $n_{\text{NGM}}$ thus agrees with the number of broken generators $n_{\text{BG}}$.

In principle, however, there is no reason not to add the following term to the effective Lagrangian \ref{phononL}, as far as it respects all assumed symmetries.
\begin{equation}
\frac{1}{2}\sum_{a,b}\rho_{ab}\partial_tu^au^b.\nonumber
\end{equation}
This term makes momentum operators non-commuting $\langle[\hat{P}_a, \hat{P}_b]\rangle=-i\rho_{ab}V$ as indicated by Eq.~\ref{charge}.  In fact, it does appear in several actual settings. The classical example is the Wigner crystal of electrons in two dimensions exposed to an external magnetic field $B_z$.  In this case $\rho_{ab}=eB_z$ as a result of the Aharonov--Bohm effect.   The phonon mode in the Wigner crystal, which is classified as type B in our scheme, has a dispersion proportional to a fractional power of $k$ due to the long-range Coulomb interaction of electrons~\cite{Fukuyama,Fractional}.

A more recent example of $\rho_{ab}\neq0$ is provided by the crystal of topological solitons, called Skyrmions, in two-dimensional ferromagnets~\cite{Skyrmion,Skyrmion2}. The Berry phase action of underlying spins gives rise to $\rho_{xy}=4\pi s n_{\text{sk}}$ where $s$ is the spin density and $n_{\text{sk}}$ is the Skyrmion number density~\cite{Stone,Nagaosa,Noncommuting}. This term is the origin of the Magnus force acting on Skyrmions~\cite{Stone} and the quadratic dispersion of the phonon in Skyrmion crystals~\cite{Nagaosa}.  When the thickness of the system is taken into account in three dimension, each Skyrmion becomes a line-like object rather than a point-like soliton and the crystalline phase is formed by a lattice of Skyrmion lines.  Low energy fluctuations of such a system were studied in Refs.~\cite{PhysRevB.90.045145,KobayashiNitta2}.

\subsubsection{Supersolids}
It is often the case in condensed matter physics that the translation symmetry is spontaneously broken together with some internal symmetries.
As a trivial setting, let us imagine two completely independent orders --- a crystalline order that breaks the continuous translation symmetry down to a lattice translation symmetry $\mathbb{R}^d\rightarrow\mathbb{Z}^d$, and another order that breaks an internal symmetry $G_{\text{int}}\rightarrow H_{\text{int}}$.  In this case phonons discussed in Sec.~\ref{subsec:crystal} and NGMs originating from the spontaneously broken internal symmetries (Sec.~\ref{sec:internal}) would simply coexist.
 
As a more nontrivial example, let us consider a system of interacting bosons with a finite density at zero temperature. When the repulsive interaction is properly designed, the Bose-Einstein condensate may develop a spontaneous crystalline order simultaneously breaking the translation symmetry and the $\text{U}(1)$ symmetry.  Such a possibility was considered in the context of the postulated supersolid phase of ${}^4\text{He}$~\cite{supersolid,supersolid2}.  Then a natural question arises: is the superfluid phonon originating from the spontaneous breaking of $\text{U}(1)$ symmetry (Sec.~\ref{subsec:superfluid}) independent from phonons associated with the crystalline order? This question was answered affirmatively by the effectively Lagrangian approach~\cite{SonPRL} and by the mean-field analysis of a Gross-Pitaevski model~\cite{Translation,KunimiKato}.

In examples we discussed so far, the straightforward extension of Eq.~\ref{countingtot} to space-time symmetries seems to work if the momentum operators are included to the definition of the matrix $\rho$ in Eq.~\ref{charge}.  An example found in Ref.~\cite{KobayashiNitta} in which the commutation relations between internal symmetries and momentum operators become nontrivial also belongs to this class.  We will see, however, that space-time symmetries are not that simple in general.

\subsection{Other spatial symmetries}
Now we move on to other spatial symmetries such as rotations and dilatations.  We still assume $t'=t$ in Eq.~\ref{transformation}.

\subsubsection{Redundancies}
In $d$-dimensional crystals, not only the translation $\mathbb{R}^d$ but also the $\text{SO}(d)$ rotation are spontaneously broken by the crystalline order. Thus the number of broken generators is at least $d$ plus $d(d-1)/2$, the latter arising from the rotation part.  Nevertheless there do not appear additional NGMs corresponding to the broken rotation symmetry in solids.  In this case, it is easy to see that a naive application of Eq.~\ref{countingtot} cannot explain this mismatch, because the matrix $\rho$ vanishes when the crystal is at rest as long as the standard algebra among the linear momentum and the angular momentum is assumed. Similar phenomenon is known to occur for spontaneous breaking of conformal symmetries even in the \emph{Lorentz-invariant} case~\cite{Salam}.  How can we account for these missing NGMs?

Low and Manohar~\cite{LowManohar} addressed this puzzle. Let $\langle\hat{\Phi}(\bm{x})\rangle$ be the order parameter detecting the breaking of generators $T_a$ (i.e., $T_a\langle\hat{\Phi}(\bm{x})\rangle\neq0$ for broken generators). According to Ref.~\cite{LowManohar}, the number of NGMs originating from spontaneous breaking of spatial symmetries is reduced by the number of nontrivial solutions to the equation
\begin{equation}
\sum_{a}c^a(\bm{x})T_a\langle\hat{\Phi}(\bm{x})\rangle=0,
\end{equation}
where $c^a(\bm{x})$ is a function of the coordinates $\bm{x}$ in the directions of unbroken translation. Using their example~\cite{LowManohar}, let us set $d=2$ and consider an order parameter that depends on $x$ but not on $y$, implying the spontaneous breaking of the momentum $P_x$ and the angular momentum $J_z$. Plugging the relation $J_z\approx xP_y-yP_x$ among these generators, one gets 
\begin{equation}
(c^{x}P_{x}+c^zJ_z)\langle\hat{\Phi}(\bm{x})\rangle=(c^{x}-c^{z}y)P_x\langle\hat{\Phi}(\bm{x})\rangle=0,
\end{equation}
for which $c^{x}(y)=yc^{z}(y)$ is a nontrivial solution. Such solutions imply redundancies among long wavelength fluctuations of the order parameter --- some Nambu--Goldstone fields $\pi$ can be eliminated in favor of a gradient of other Nambu--Goldstone fields $\partial_\mu \pi$~\cite{LowManohar}. This pattern of symmetry breaking, forming a crystalline order in one direction that also implies a breaking of associated rotation symmetry, occurs in smectic phases of liquid crystals, and the reduction of low-energy modes in this context has been extensively studied (see e.g.~\cite{CL,Prost,Grinstein1,Grinstein2}).  More generally, the reduction of independent degrees of freedom for space-time symmetries through this mechanism is called the inverse-Higgs effect~\cite{LowManohar,Ivanov1975,InverseHiggs,Noumi,Rothstein2018}.


This, however, fails to explain the situations which there is no unbroken continuous translation symmetries.  Also, the relation among generators should be formulated more rigorously in term of conserved charge densities in Eq.~\ref{charge0}.  These points were refined in Ref.~\cite{Redundancies}.  The number of independent Nambu--Goldstone fluctuations is reduced by one for every linear combination of Noether densities of globally conserved charges  that annihilates the ground state:
\begin{equation}
\sum_{a}c^a(\bm{x})\hat{q}_a(\bm{x})|\text{GS}\rangle=0.\label{NoetherC}
\end{equation}
Here, coefficients $c^a(\bm{x})$ are allowed to depend on all coordinates.  For example, in a 2D crystal that breaks $\hat{P}_i=\int d^2x \hat{q}_i$ ($i=x,y$) and $\hat{J}_z=\int d^2x \hat{q}_z$,  the exact relation among the conserved charge densities $\hat{q}_z=x\hat{q}_y-y\hat{q}_x$ implies that $(\hat{q}_z-x\hat{q}_y+y\hat{q}_x)|\text{GS}\rangle=0$. This is an example of Eq.~\ref{NoetherC} with $c^z=1$, $c^x=y$, and $c^y=-x$.  A broken generator $\hat{Q}_a$ produces an independent NGM when it is not involved in any linear combination in Eq.~\ref{NoetherC}.  The counting of the independent low-energy fluctuations applicable to a finite temperature was formulated in Ref.~\cite{Hayata}.

\subsubsection{Overdamping}
In the previous section, we explained a general criterion for the appearance of independent NGMs originating from spontaneously broken space-time symmetries.  Here, we assume they do exist and examine their low-energy properties.

In general, ordered phases contain many dynamical degrees of freedom apart from NGMs. If other excitations are all gapped, they never affect the low-energy physics in any essential manner --- one can safely ``integrate out" gapped modes that only renormalize parameters.  On the other hand, if there exists other gapless modes, they can, in principle, destroy NGMs via interactions.  However, the microscopic symmetry of the system encoded in the effective Lagrangian puts severe restrictions on the way NGMs interact among themselves and with other degrees of freedom~\cite{Weinberg}.  As a result, NGMs are robust and remain well-defined propagating modes at least in the low-energy, long-wavelength limit.  This is consistent with the statement of the Nambu--Goldstone theorem --- the very fact that it could predict the appearance of massless particles without caring about other degrees of freedom  already implies the stability of NGMs.  Below we argue that such a nice low-energy property is absent in the case of NGMs originating from spontaneously broken spatial symmetries, except for the ordinary phonons.  

To understand the key point through an educative example, let us review the nematic Fermi fluid discussed in Refs.~\cite{Oganesyan,FradkinAR}. In this example, a circular Fermi surface of spinless fermions in $(2+1)$ dimensions spontaneously deforms into an elliptic shape, breaking the continuous rotation symmetry down to a discrete two-fold rotation. As the translation symmetry remains unbroken, there is no linear combination that satisfies Eq.~\ref{NoetherC} and a Nambu--Goldstone field $\theta(\bm{x},t)$ can be introduced. It transforms nonlinearly under the rotation by an angle $\epsilon$, 
\begin{equation}
\theta'(R\bm{x},t)=\theta(\bm{x},t)+\epsilon,
\label{theta2}
\end{equation}
where $R=e^{-i\epsilon\sigma_y}\in \text{SO}(2)$ represents the rotation matrix. The fermion field $\psi(\bm{x},t)$ is a scaler $\psi'(R\bm{x},t)=\psi(\bm{x},t)$.   Using these ingredients, one can write down the following rotation-symmetric Lagrangian
\begin{align}
\mathcal{L}_{\text{eff}}&=\Big[\psi^*(i\partial_t+\mu)\psi-\frac{1}{2m}\bm{\nabla}\psi^*\cdot\bm{\nabla}\psi\Big]+\Big[\frac{1}{2}\bar{g}(\partial_t{\theta})^2-\frac{1}{2}g(\bm{\nabla}\theta)^2\Big]\nonumber\\
&+\lambda\big[(\partial_x\psi^*\partial_x\psi-\partial_y\psi^*\partial_y\psi)\cos2\theta+(\partial_x\psi^*\partial_y\psi+\partial_y\psi^*\partial_x\psi)\sin2\theta\big],\label{nematic}
\end{align}
which can be derived by a mean-field approximation of a spinless electron model~\cite{Oganesyan}.  The first term represents the original circular Fermi sea, the second one gives the bare linearly dispersive NGM, and the third one  describe the interaction between them.

Assuming that the fluctuation $\theta$ is small, we may Taylor expand the interaction in the power series of $\theta$. The zero-th order term together with the first term produces an elliptic Fermi surface, i.e., the set of $(k_x,k_y)$'s satisfying $\mu=(\frac{1}{2m}-\lambda)k_x^2+(\frac{1}{2m}+\lambda)k_y^2$.  Surprisingly, the interaction term linear in $\theta$ [$2\lambda (k_x'k_y+k_y'k_x)\psi_{\bm{k}}^*\psi_{\bm{k}}\theta_{\bm{q}}$ in the Fourier space] does not contain any derivative acting on $\theta$.  The coefficient of this term becomes $4\lambda k_xk_y$ in the limit of $q\rightarrow0$, which does not vanish at most of parts on the Fermi surface.   A perturbative calculation with this non-derivative coupling signals the breakdown of Fermi-liquid theory and overdamping of the NGM~\cite{Oganesyan}.  Namely, the NGM, after dressed up with particle/hole excitations near the distorted Fermi surface, looses its particle nature.  

The transformation of the $\theta$-field in Eq.~\ref{theta2} is almost identical to that of superfluids in Eq.~\ref{theta}. The crucial difference lies in the argument of $\theta'$ ($R\bm{x}$ versus $\bm{x}$), which distinguishes the spatial rotation (a space-time symmetry) from the phase rotation (an internal symmetry).  This difference results in the presence/absence of the ``non-derivative" interaction of NGMs we discussed just now.

This scenario is not restricted to the above specific setting and the generalization to wider class of space-time symmetries was performed in Ref.~\cite{NFLPNAS}. According to the study, a NGM will be overdamped in the presence of Fermi surface if the NGM originates from spontaneous breaking of a generator $\hat{Q}_a$ that does not commute with the momentum operator, i.e.,
\begin{equation}
[\hat{Q}_a,\hat{\bm{P}}]\neq0.\label{criterion}
\end{equation}
This is actually the case for almost all of space-time symmetries other than the translation symmetry.  Examples of such situations have been discussed in Refs.~\cite{Ruhman,Yasaman,Eduardo1}.  

To summarize this section, for spontaneously broken spatial symmetries, one has to pay attention to two additional subtleties: (i) The number of independent fluctuations may be fewer than the number of broken generators. Namely, introducing one field for every broken generators may be redundant. This mismatch cannot be captured by naively extending the counting rule for internal symmetries [Eq.~\ref{countingtot}] but can be seen, instead, by listing up relations of the form of Eq.~\ref{NoetherC}.  (ii) Even when additional low-energy degrees originate from spontaneous breaking of spatial symmetries, some of them may not form a propagating mode, being overdamped via interactions with other low-energy degrees. In the presence of a Fermi surface in the ordered phase, Eq.~\ref{criterion} provides the criterion for overdamping.

\section{When does symmetry breaking occur?}
\label{sec:LSM}
The Nambu--Goldstone theorem \emph{assumes} spontaneous breaking of a global symmetry and predicts the consequence. In this section, we address two complementary questions. 
(i) Can any symmetry, in principle, be spontaneously broken if the model Hamiltonian is properly designed? 
(ii) For symmetries that can indeed be broken in some setting, when do they tend to be broken?

\subsection{Time translation symmetry}
When $G$ is an internal symmetry, the answer to the first question always seems to be positive. Let $\langle\hat{\Phi}_\alpha(\bm{x})\rangle$ be an order parameter transforming nontrivially under $G$ as in Eq.~\ref{transformation}.  
By modifying the Hamiltonian $\hat{H}$ to $\hat{H}-\lambda \int d^3x\sum_\alpha\hat{\Phi}_\alpha(\bm{x})^\dagger\hat{\Phi}_\alpha(\bm{x})$ with a large enough coefficient $\lambda$,  the order parameter $\langle\hat{\Phi}_\alpha(\bm{x})\rangle$ would be non-vanishing in lowest energy states and $G$ would be spontaneously broken.

Such a naive argument does not apply to the time translation symmetry $t\rightarrow t'=t+\epsilon$ generated by the Hamiltonian itself.
In 2012, Wilczek~\cite{Wilczek} proposed a possibility of a new phase, dubbed ``quantum time crystal," in which the quantum ground state spontaneously breaks the time translation symmetry down to a discrete subgroup. If such a phase existed, it would be a temporal analog of ordinary crystals discussed in Sec.~\ref{subsec:crystal}.  However, several follow-up studies~\cite{Bruno,Absence} clarified that the time-translation symmetry cannot be spontaneously broken in any ground state or thermal equilibrium.  The time translation symmetry is the single known example of ``never-be-broken" global symmetries, and it would be an interesting future work to pin down the general criterion for space-time symmetries of this type.

Remarkably, Wilczek's proposal eventually led to the recent discovery of ``discrete time crystals"~\cite{Khemani,Else,Yao,TC1,TC2,Sacha} in non-equilibrium, externally driven systems. This exciting topic will be covered in the review article by Nayak (ARCMP vol. 11).  A non-equilibrium time crystal may support diffusive NGMs, which has recently been studied in Refs.~\cite{Hongo2019,HayataHidaka2}. See also Ref.~\cite{Sieberer_2016} for spontaneous breaking of more standard symmetries in quantum open systems.

\subsection{Lieb-Schultz-Mattis theorem}
The Lieb-Schultz-Mattis theorem is a theorem imposing a general, non-perturbative constraint on the low-energy spectrum of many-body quantum systems in arbitrary spatial dimension $d$.
It assumes two symmetries: a $\text{U}(1)$ symmetry that defines an \emph{integer-valued} charge $\hat{Q}$ via the Noether theorem (Sec.~\ref{sec:basics}) and a discrete translation symmetry that defines a fundamental domain, called ``unit cell." (In other to make $\hat{Q}$ integer-valued, one should properly choose the overall factor and the offset in the definition of $\hat{Q}$).  These symmetries can be just a subgroup of the full symmetry group of the system.  
Given them, \emph{the average U(1) charge per unit cell in a ground state} is well-defined even in the thermodynamic limit.  We call this quantity the ``filling" $\nu$, borrowing the terminology used in electronic systems. The Lieb-Schultz-Mattis theorem states that it is possible to isolate a unique ground state in energy from other states only when the filling $\nu$ is an integer. In other words, if $\nu$ is not an integer, either ground state degeneracy or gapless excitations must exist. 

This theorem was originally formulated for Heisenberg spin models (Sec.~\ref{subsec:Heisenberg}) in $1+1$ dimensions~\cite{Lieb1961}.  For this class of models, the theorem may be viewed as a proof of the half of the Haldane conjecture~\cite{Affleck1986}, since it predicts gapless excitations for half-integer spin chains, assuming that the ground state is symmetric. Later the theorem was generalized to a wider class of models with two symmetries stated above in arbitrary spatial dimensions~\cite{Yamanaka,Oshikawa2000,Hastings2004}.  Recently the theorem has been refined under an assumption of crystalline symmetries~\cite{Sid,LSMPNAS,LSMMSG}.

A ground state degeneracy often occurs as a result of spontaneous symmetry breaking.  In $d\geq2$, there is a more exotic possibility called ``topological" degeneracy~\cite{XGW}, which has been the subject of a large number of recent studies as it features many intriguing phenomena including ``charge fractionalization"~\cite{PhysRevLett.96.060601}.  A gapless excitation does not have to be NGMs either. Therefore we cannot say anything deterministic from this theorem but we can say at least ``something" needs to occur.

As an application, let us consider a system of bosons with a nonzero density in free space.  Since the continuous translation symmetry includes arbitrary discrete translation subgroups, one can assume a unit cell of any size, and the filling $\nu$ can take any positive value. Therefore, a unique ground state with an excitation gap is prohibited.  This explains why the Bose-Einstein condensation (Sec.~\ref{subsec:superfluid}), exhibiting both a ground state degeneracy and gapless excitations, is such common for bosonic systems at finite density at zero temperature.

\section{Concluding remarks}
In this review we focused on NGMs originating from spontaneous breaking of global symmetries.  Formulae for  internal symmetries presented in Sec.~\ref{sec:internal} do not assume the Lorentz invariance and thus are applicable to both relativistic and nonrelativistic systems.  Subtleties of NGMs associated with space-time symmetries are explained in Sec.~\ref{sec:spacetime}.  Two relevant issues, which have never been covered in existing reviews on this subject, were addressed in Sec.~\ref{sec:LSM}.

Before closing, let us briefly mention important related topics we could not discuss in this review.  At the tree-level approximation, the number of flat directions of order parameter fluctuations can be accidentally larger than the minimum number protected by spontaneous symmetry breaking. In that case one encounters \emph{quasi-}Nambu--Goldstone modes (or \emph{pseudo-}Nambu--Goldstone modes)~\cite{PhysRevLett.29.1698}. See, for example, Refs.~\cite{UchinoUeda,Ueda1} for their realization in cold atom systems and Ref.~\cite{NittaTakahashi} for the counting rule of quasi-NGMs.  When spontaneously broken symmetries are \emph{fermionic}, the resulting low-energy, long-wavelength excitations are also fermionic and are called Nambu--Goldstone fermions~\cite{VOLKOV1973109,SALAM1974465}. Refs.~\cite{PhysRevLett.100.090404,PhysRevA.92.063629,PhysRevD.94.045014} investigated Nambu--Goldstone fermions in nonrelativistic systems with a supersymmetry.  When global symmetries are ``gauged" and promoted to local ones, the Higgs mechanism may eliminate NGMs and produce a mass to gauge fields~\cite{PhysRevLett.13.508,PhysRevLett.13.321}.  The Higgs mechanism in a nonrelativistic setting has been studied in Refs.~\cite{PhysRevD.83.125009,Gauge,Gongyo}. Finally, the notion of global symmetries has been generalized to ``$q$-form" symmetries~\cite{Gaiotto2015}, where $q=0$ corresponds to the conventional class discussed in this review.  In this scheme, photons in electrodynamics are viewed as a type-A NGM associated with a one-form symmetry~\cite{Ethan}, and there are also examples of type-B NGMs~\cite{PhysRevD.93.085036,Ozaki2017,1903.02846}. Developing an extended formalism describing these NGMs originating from spontaneously broken higher-form symmetries will be an interesting future work.

As we have seen, spontaneous symmetry breaking is quite ubiquitous and examples can be found in almost all areas of physics.  It is certainly not possible to exhaust all related topics and we stop here.
We hope this review helps readers with various background to get into this old but everlasting subject.

\section*{ACKNOWLEDGMENTS}
This review is based on what the author learned through collaborations with Tom\'{a}\v{s} Brauner, Hitoshi Murayama, Ashvin Vishwanath, and Masaki Oshikawa over years.


%

\noindent

\bibliography{reference}

\end{document}